\documentclass[seceq]{ptptex}

\usepackage{graphicx}
\usepackage{epsfig}
\usepackage{amssymb}

\title{
Compatibility of Exotic States with Neutron Star Observation
}

\author{
Chang Ho \textsc{Hyun}\footnote{ e-mail address: hch@meson.skku.ac.kr}
}

\inst{
Institute of Basic Science, Sungkyunkwan University, 
Suwon 440-746, Korea\\
and \\
Department of Physics, Seoul National University,
Seoul 151-742, Korea
}

\abst{
We consider the effect of hard core repulsion in the baryon-baryon
interaction at short distance to the properties of a neutron star. 
We obtain that, even with hyperons in the interior of a neutron star, 
the neutron star mass can be as large as $\sim 2 M_\odot$.
}

\begin{document}

\maketitle

\section{Introduction}

Reports on recent observations of pulsars in various binary systems
show that the maximum mass of a neutron star can be large as 
$(1.7 \sim 2.1) M_\odot$ \cite{nice,review1}.
Most of them still have large uncertainty, 
but a few are within the above range with relatively small error bars.
The possibility of large mass of a neutron star thus 
has led to a claim that exotic states of matter at high densities
are not necessary
in the neutron star as far as its mass is concerned \cite{ozel},
but there also appeared a counter argument that the large mass
does not necessarily rule out the exotic states \cite{klahn}.
There are many sources of uncertainties at high densities, {\it e.g.}
state of matter, constituent particles and their interactions,
but the information available to reduce the uncertainties is not 
sufficient yet.

We revisit the neutron star mass problem with a simple
phenomenological approach.
One fixed point of nuclear matter physics is the nuclear saturation density;
its properties such as density, binding energy, symmetry energy,
and compression modulus are fairly well constrained. 
We describe these saturation properties in terms of quantum hadrodynamics
(QHD) \cite{qhd}.
The other fixed point we choose is the hard core repulsion at short range.
Though it is a kind of artifact adopted to describe the nucleon-nucleon
data, its role is clear in many phenomena of nuclear physics.
The effect of hard core can be parametrized with an excluded volume
in the estimation of thermodynamic variables 
\cite{rischke,kagiyama,cleymans}.
It has been more frequently employed to explain the phase transition
in the relativistic heavy ion collision environment, 
and could describe well the transition
from hadronic to quark-gluon plasma phase \cite{epjc02}.

In this work, 
by including hard cores in the interaction of baryons, we explore 
the bulk properties of the neutron star, and compare the result with 
the recent observation of heavy neutron star masses.
This paper is outlined as follows. In the next section, we briefly
address the basic formalism of QHD with hard core.
The next section comes up with numerical results, and
brief concluding remarks are drawn in the following section.

\section{Formalism}

We employ the QHD Lagrangian,
\begin{eqnarray}
{\cal L} &=& \sum_B \bar{\psi}_B \left(
i \partial \cdot \gamma - m_B + g_{\sigma B} \sigma 
- g_{\omega B} \gamma_0 \omega_0 
- g_{\rho B} \tau_3 \gamma_0 b_{30} \right) \psi_B \nonumber \\
& & - \frac{1}{2} m^2_\sigma \sigma^2 + \frac{1}{2} m^2_\omega \omega^2_0
+ \frac{1}{2} m^2_\rho b^2_{30} 
- \frac{b}{3} m_N (g_{\sigma N} \sigma)^3 
- \frac{c}{4} (g_{\sigma N} \sigma)^4 \nonumber \\
& &+\sum_{l = e,\, \mu} \bar{\psi}_l (i \partial \cdot \gamma - m_l) \psi_l,
\end{eqnarray}
where the baryon species $B$ includes octet baryons, and 
$\sigma$, $\omega_0$ and $b_{30}$ are non-vanishing meson fields
in the mean field approximation.
When we account for the forbidden region due to hard core,
the baryon density is redefined as
\begin{equation}
\rho = \frac{\rho'}{1 + v_{\rm ev} \rho'},
\end{equation}
where $\rho'$ is the density in the case of point particle
and $v_{\rm ev}$ the excluded volume.
We assume $v_{\rm ev} = \frac{4}{3}\pi r^3_0$ where $r_0$ is the 
radius of hard core, which is treated as a free parameter in
our consideration.
Consistency with thermodynamic relations and self-consistency conditions
alter the form of state variables (pressure, chemical potential, 
energy density and etc) and equation of motion of $\sigma$-meson field 
from those of point particle ones.
The explicit formulas and equations can be found in old 
\cite{rischke, kagiyama, cleymans} and recent \cite{panda, aguirre}
publications. 

Three meson-nucleon coupling constants $g_{\sigma N}$, $g_{\omega N}$
and $g_{\rho N}$ and two $\sigma$-meson self interaction coefficients
$b$ and $c$ are fitted to five saturation properties,
the saturation density (0.17 fm$^{-3}$), binding energy (16.0 MeV),
symmetry energy (32.5 MeV), compression modulus (300 MeV) and
nucleon effective mass ($0.75 m^*_N$), with a given hard core radius $r_0$.
Meson-hyperon coupling constants are determined by quark counting rules,
$g_{MY} = g_{MN} \sum_{q=u,d} n_{qY}/3$, 
where $g_{MY}$ is the meson-hyperon coupling constant,
$n_{qY}$ is the number of $u$ and $d$ quarks in a hyperon species $Y$ 
and $g_{MN}$ is the meson-nucleon coupling constant.
As for the hard core radius of hyperons, we assume the same value as
that of the nucleon for simplicity.

Table~\ref{tab:couplings} summarizes the parameters determined
from the given saturation properties and hard core radii.
\begin{table}[tbh]
\begin{center}
\begin{tabular}{c|c|c|c|c|c}\hline
$r_0$ (fm) & $g_{\sigma N}$ & $g_{\omega N}$ &
$g_{\rho N}$ & $b\, (\times 10^3)$ & $c\, (\times 10^3)$ \\ \hline
0   & 8.44 & 8.92 & 7.76 & 3.97 & 4.00 \\ 
0.2 & 8.43 & 8.91 & 7.72 & 3.80 & 4.37 \\
0.3 & 8.39 & 8.89 & 7.64 & 3.38 & 5.26 \\
0.4 & 8.30 & 8.85 & 7.47 & 2.51 & 7.11 \\
0.5 & 8.16 & 8.78 & 7.19 & 0.93 & 10.47 \\ \hline
\end{tabular}
\end{center}
\caption{Meson-nucleon coupling constants and coefficients $b$ and $c$
fitted to a set of saturation properties described in the text
with a given $r_0$ value.}
\label{tab:couplings}
\end{table}

\section{Numerical result}

\begin{figure}[tbh]
\begin{center}
\epsfig{file=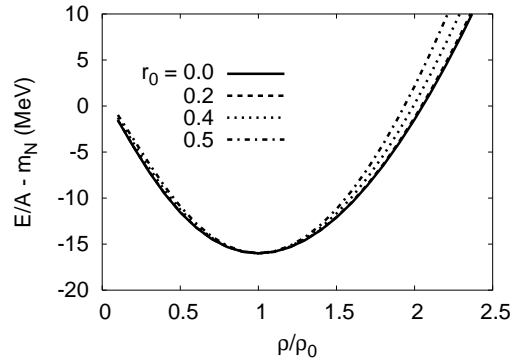, width=7cm} 
\end{center}
\caption{Binding energy of a nucleon in the 
symmetric nuclear matter with different hard core radii.}
\label{fig:eos}
\end{figure}

Fig.~\ref{fig:eos} shows the binding energy per a nucleon in the
symmetric nuclear matter with different hard core radii. 
Though the saturation properties are the same regardless of $r_0$ values, 
the equation of state becomes stiffer
at high densities with a larger $r_0$ value.
%

The equation of state of neutron star matter is determined 
self-consistently by the baryon number conservation, 
charge neutrality, $\beta$-equilibrium of baryons and leptons, 
and equations of motion of meson fields.
Once the equation of state is determined, the mass-radius relation
of a neutron star
can be obtained by solving Tolman-Oppenheimer-Volkoff (TOV) equation.
Table~\ref{tab:massradius} shows the maximum mass of a neutron star,
corresponding radius and central density 
with nucleons only (columns of ``$np$") and 
with hyperons (columns of ``$npY$").
Consistent with the behavior of the equation of state at high densities
in Fig.~\ref{fig:eos},
the maximum mass becomes larger with a larger $r_0$ value.
For $np$ case, when $r_0=0.5$ fm, the increase amounts to 10\% 
of the maximum mass without hard core. 
With hyperons, the maximum mass increases
to the range of large mass in recent observations when
$r_0 \gtrsim 0.3$ fm. 
%
\begin{table}[tbh]
\begin{center}
\begin{tabular}{c|c|c|c||c|c|c}\hline
 & \multicolumn{3}{c||}{$n\, p$} & \multicolumn{3}{c}{$n\, p\, Y$} \\ \hline
$r_0$ (fm) & $M$ ($M_\odot$) & $R$ (km) & $\rho_{\rm cent}$ ($\rho_0$) &
$M$ ($M_\odot$) & $R$ (km) & $\rho_{\rm cent}$ ($\rho_0$) \\ \hline
0   & 2.10 & 10.9 & 6.4 & 1.53 & 11.3 & 6.1 \\
0.2 &  -   &  -   &  -  & 1.58 & 11.4 & 6.1 \\
0.3 & 2.14 & 11.1 & 6.2 & 1.70 & 11.5 & 5.9 \\
0.4 & 2.20 & 11.4 & 5.9 & 1.97 & 12.2 & 5.2 \\
0.5 & 2.34 & 11.7 & 5.4 &  -   &  -   &  -  \\ \hline
\end{tabular}
\end{center}
\caption{Maximum mass $M$ in units of solar mass, and corresponding
radius $R$ in km and central density $\rho_{\rm cent}$ in unit of 
the saturation density $\rho_0$. }
\label{tab:massradius}
\end{table}

\section{Conclusion}

We investigated the maximum mass of a neutron star
in a simple phenomenological approach where the hard-core repulsion
is included in the QHD model.
The hard core radius is treated as a free parameter,
and the meson-nucleon coupling constants are fixed
identical saturation properties.
We obtained the equation of state of neutron star matter
that satisfies thermodynamic equations and 
self-consistency conditions.
Solving TOV equation, we obtained the mass-radius 
relation of a neutron star.
Our result shows that the maximum mass with hyperons 
can be as large as observed masses with a hard core radius
$r_0 \gtrsim 0.3$ fm.
These values of $r_0$ are in the range of hard core radius
$0.3 \sim 0.6$ fm in well-known hard-core potential models such as 
Hamada-Johnston \cite{hamada} or Reid \cite{reid}.
More investigations are necessary to figure out the uncertainties.
For instance, the hard core size of hyperons can matter.
The effect of hard cores to the formation of
other exotic states such as meson condensation or 
deconfined quark phases is also worthy to be studied.

To conclude, the our result shows that the hyperon matter, which is
known to give the biggest effect to the mass-radius relation of 
a neutron star among possible exotic states in the interior of
a neutron star, is not necessarily incompatible with the observed mass.

\section*{Acknowledgments}
The author thanks the Yukawa Institute for Theoretical Physics 
at Kyoto University, where this work was initiated and developed 
during the YKIS2006 on "New Frontiers on QCD". 
Author is grateful to Shung-ichi Ando for reading the manuscript.
This works was supported by the Basic Research Program of
Korea Science \& Engineering Foundation (R01-2005-000-10050-0).

\end{document}